\documentclass[showpacs,superbib,10pt,preprint,final,twocolumn,aps]{revtex4}
\usepackage{amsfonts}
\usepackage{amsmath}
\usepackage{amssymb}
\usepackage{graphicx}%
\begin{document}
\input epsf.sty \flushbottom
\title{Transport in nanoscale systems: the microcanonical versus grand-canonical picture}
\author{M. Di Ventra\cite{MD}}
\affiliation{Department of Physics, University of California, San Diego, La Jolla, CA 92093-0319}
\author{T. N. Todorov\cite{TNT}}
\affiliation{School of Mathematics and Physics, Queen's University of Belfast,
Belfast BT7 1NN, United Kingdom}

\begin{abstract}
We analyse a picture of transport in which two large but finite charged electrodes discharge  
across a nanoscale junction. We identify a functional whose minimisation, within the space 
of all bound many-body wavefunctions, defines an instantaneous steady state. 
We also discuss factors that favour the onset of steady-state conduction in such systems, make a connection with  
the notion of entropy, and suggest a novel source of steady-state noise. 
Finally, we prove that the true many-body total current in this closed system is given exactly 
by the one-electron total current, obtained from time-dependent density-functional theory. 
\end{abstract}
\maketitle
When a metallic nanojunction between two macroscopic
electrodes is connected to a battery, electrical current 
flows across it~\cite{prec}. The battery provides, and maintains, the charge 
imbalance between the electrode surfaces, needed to sustain steady-state conduction
in the junction. This static non-equilibrium problem is usually described according to the 
Landauer picture~\cite{landauer57}. In this picture, the junction is connected to a 
pair of defect-free metallic leads, each of which is connected to its own distant 
infinite heat-particle reservoir. The pair of reservoirs represents the battery. 
Each reservoir injects electrons into 
its respective lead with the electrochemical potential appropriate to the bulk of that reservoir. 
Each injected electron then travels undisturbed down the respective lead to the junction,
where it is scattered and is transmitted, with a finite probability, into the other lead.
From there it flows, without further disturbance, into the other reservoir. The reservoirs are conceptual 
constructs which allow us to map the transport problem onto a truly stationary scattering one,
in which the time derivative of the total current, and of all other local
physical properties of the system, is zero. By doing so, however, we arbitrarily  
enforce a specific steady state whose microscopic nature is not, in reality, known {\em a priori}. 
The Landauer construct is highly plausible in the case of non-interacting electrons.
In the case of interacting electrons, however, it is not at all obvious that 
the steady state in the Landauer picture is the same as that which would be 
established {\em dynamically} by the electrons originating from the battery and 
flowing across the junction. It is also not obvious whether the same steady state can be 
reached with different initial conditions.

There is, however, an alternative picture of DC conduction. In that picture, 
we dispense with the battery and we think of the current as a long-lived, but ultimately 
transient, discharge of a macroscopic, but finite, capacitor \cite{sankey,tod01,horsfield}.
This view has great appeal. We now have a finite system, with a finite
number of electrons and nuclei. This system can be visualised and realised practically.
It can be described dynamically, at least in principle, by solving a finite, closed set of
equations of motion for the particles in the system. In this picture, 
the system is allowed to find its own electronic structure during the discharge, without
the imposition of {\em a priori} assumptions about what this electronic 
structure should look like. This ``microcanonical'' picture, in which the 
conventional notion of transport as an open-boundary, ``grand-canonical'' 
problem is replaced by the idea of a long-lived discharge of an isolated finite system,
is represented schematically in figure 1.
In keeping with the microcanonical framework that we are interested in, we do not introduce 
dissipative effects in this system~\cite{horsfield,car,baer}.

\begin{figure}
\includegraphics[width=.40\textwidth]{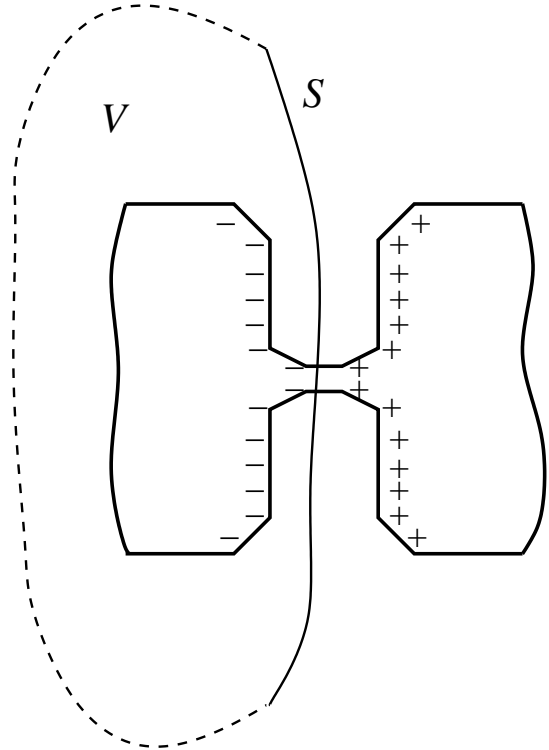}
\caption{A nanoscale junction between two large 
but finite metallic charged electrodes. The details are discussed in the text.}
\label{fig1}
\end{figure}

The purpose of this paper is twofold. First, we seek to establish
a mapping of the conventional steady-state 
transport problem onto the present microcanonical one.
To this end, we define in variational terms a class of dynamical states
for the closed geometry in figure 1 that we call 
instantaneous global quasi steady states. 
We then give arguments for their robustness. 
Second, we demonstrate that the exact total current 
(as opposed to current density) in the true
interacting many-electron system is identical to
the one-electron current, obtained from 
time-dependent density-functional theory (TDDFT)~\cite{RungeGross}.
The closed, finite nature of the system 
is a requirement for this demonstration. The result is valid
throughout the time evolution of the electronic system,
regardless of whether or not the system gets anywhere near a steady state.
This result places on a rigorous footing the application of 
time-dependent one-electron methods to transport in nanoscale systems.

Our motivation is the tremendous physical
and computational appeal of the resultant microcanonical
picture of conduction: it gives us, in principle, a formally exact,
dynamical description of many-electron transport within a one-electron picture. 
It does so, furthermore, in a way that eliminates the numerically cumbersome 
implementation of scattering boundary conditions and the unphysical
notion of infinite systems that continually plague one in the 
traditional static approach to transport.

Before discussing the closed system in figure 1, let us return to the transport problem 
in the open-boundary approach. In the self-consistent steady state, a net
electron current from one electrode into another
across a nanojunction is accompanied by an excess of 
electrons in one electrode and a deficit of electrons 
in the other~\cite{MDLang}. These charges take the form of 
surface charge densities, present within a screening length
of the electrode surface. Thus, steady-state conduction goes hand in hand
with surface charges on each side of the junction: a negative sheet of charge on one side and 
a positive sheet on the other, as if two infinite capacitor plates were 
present. The transport problem can then be viewed as a continuous attempt by the electrons to 
``passivate'' the surface charges on each side of the junction. In other words, 
the steady-state transport problem in the traditional open-boundary approach
{\em is}, effectively, the continuous discharge of an {\em infinite} capacitor.

In real life there is no such thing as an infinite capacitor.
We must therefore consider the discharge of a finite, though possibly very large,
capacitor. If $C$ is its capacitance
and $R$ is the resistance across which it discharges, then,
according to classical circuit theory,
the discharge will take place over a characteristic time $RC$. 
The larger $C$, for a given $R$, the
more stationary the properties of the system will appear, in a temporally
local sense, at any one stage of the discharge. This ultimately transient,
but very slowly varying, conducting state indeed encapsulates the intuitive picture 
of a DC steady state that we all have.

There are stardard {\em time-dependent} approaches that 
allow the DC steady state, in the limit of infinite system size, 
to be described from a microscopic point of view.
One such approach is linear response theory.
Here, we imagine that the electrons in the electrode-junction-electrode 
system start off in the ground state. Then, 
at some initial time $t=0$, a static external electric field (which need
not be uniform) is applied, and the electronic response for $t>0$ is calculated
to first order in the applied field. If we take the size of the 
electrodes to infinity (at least in the longitudinal direction) 
and then take the limit $t\rightarrow \infty$, 
with the further assumption that the electrons are non-interacting, 
we may follow the arguments of Stone and Szafer \cite{saferstone} to
obtain the two-probe one-electron 
Landauer formula $G=(2e^2/h)\,Tr\{{\bf t}{\bf t}^{\dagger}\}$, 
where $G$ is the conductance and ${\bf t}$ is the transmission matrix. 
The two-probe Landauer formula can therefore be thought of as describing
the steady state of a system of non-interacting electrons flowing between two 
infinitely large electrodes, under an external field. 
However, this approach is of no use to us
here because it explicitly requires the system size
to be taken to infinity, which is specifically what
we do not wish to do. We observe also that this approach does not take account
of electron-electron interactions. 
Of course, formally, one may write down the linear response calculation
in many-body form, but the result is in general intractable. If, on the other hand,
we choose the one-electron route of TDDFT, then in the linear response calculation,
on top of the external field, we must include the additional time-varying
one-electron potential due to the dynamical evolution of the electron density itself.
It then stops being obvious whether, and in what form, the one-particle
Landauer formula would survive. While the possible resultant 
corrections to the open-boundary Landauer formula constitute a very 
interesting line of work, the above approach to the steady state, once again,
falls beyond our goals here, because of our specific concern with 
{\em finite} systems.

An alternative dynamical approach, in a minimal discrete real-space basis set \cite{tod01}, 
would be to start with two charged but electronically
decoupled electrodes, connect them by a junction and follow the ensuing 
discharge \cite{wingreen}.
If once again we take the system size to infinity and then take the limit
$t\rightarrow \infty$, while treating the electrons as 
non-interacting, then we would end up solving the usual stationary 
Lippmann-Schwinger equation for the one-electron wavefunctions
in an open-boundary system \cite{tnt,MDLang}. 
This approach would once again result in the 
standard one-electron transport formulae \cite{wingreen,tod01}. However, 
this approach does not help us here for the same reason as above:
it relies on taking the system size to infinity. By analogy with the earlier
remarks, we note that if we include electron-electron interactions
in the {\em time-dependent} calculation in the form of an additional 
dynamical one-electron potential that depends on the time-evolving electron density, then
it is not obvious that the resultant infinite-system, long-time behaviour
would be the same as the self-consistent solution in the standard
{\em time-independent} one-electron scattering open-boundary approach, in which one solves
the one-electron static Lippmann-Schwinger equation iteratively
and self-consistently, with a given functional relation 
between density and one-electron potential, and with given fixed
incoming one-electron distribution functions \cite{tod01}.

We now return to our finite isolated 
system in figure 1. We release the electrons from an arbitrary but definite
initial state, characterised by a charge imbalance between the
electrodes. We thenceforth allow the electrons to propagate dynamically.
After an initial transient time (related to the initial state of the
electrons and to the electron-electron relaxation time inside 
the capacitor plates) during which electrons and holes first start
to traverse the nanojunction from opposite sides, we expect a 
quasi steady state to be established. We expect it to persist
until the time when multiple electron reflections off the far boundaries of the system
begin to develop. From then on, the electronic system will 
oscillate in time among several many-body states forever, if we neglect 
dissipative effects, as we do here \cite{comment0}. 
We are interested in the intermediate quasi steady state.

Our first task is to define this quasi steady state
in variational terms. By appealing to the intuitive concept of steady state in the case of
a macroscopic classical capacitor, considered earlier, we adopt the view
that a steady state is one in which the temporal variation of local
properties is minimal. A measure of temporal variations is provided by 
the functional
\begin{equation}
A[\rho] = \int_{t_{1}}^{t_{2}}\,dt\,\int_{{\rm all\,space}}\,d{\bf r}\, 
\left( \frac{\partial \rho({\bf r},t)}{\partial t}\right)^{2}
\label{a}
\end{equation}
where $\rho({\bf r},t)$ is the density of the electron gas in the
system depicted in figure 1 and $(t_{1},t_{2})$ is some time interval of interest.
Let us first perform an unconstrained variational minimisation of $A$ with respect to $\rho$, for a given 
$\rho({\bf r},t_{1})$ and $\rho({\bf r},t_{2})$. The result is
\begin{equation}
\frac{\partial^{2} \rho({\bf r},t)}{\partial t^{2}}=0\,\,\,\,\,\,\forall\,{\bf r}\,,\,\,\,\,\,
\forall\,t\in (t_{1},t_{2})
\label{rhodd}
\end{equation}
To interpret this result, let us consider the total current, $I_{S}$, through
an open surface $S$ across the electrode-junction-electrode system,
as shown in figure 1:
\begin{equation}
I_{S}=I_{S}(t) = \int_{S} {\bf j}({\bf r},t)\cdot d{\bf S}
\label{is}
\end{equation}
where ${\bf j}({\bf r},t)$ is the current density of the electrons.
We close $S$ in the vacuum, as is indicated by
the dashed part of the curve in figure 1, sufficiently far from the boundaries of
the system to enable us to ignore any contributions
to the surface integral over the dashed part of the curve \cite{comment1}.
By using the continuity equation
\begin{equation}
{\bf \nabla}\cdot {\bf j}({\bf r},t) + \frac{\partial \rho({\bf r},t)}{\partial t} = 0
\label{cont}
\end{equation}
and by invoking Gauss's theorem, we find
\begin{equation}
\frac{dI_{S}(t)}{dt} = -\int_{V} \,d{\bf r} \,\frac{\partial^{2}\rho({\bf r},t)}{\partial t^{2}}
\label{didt}
\end{equation}
where $V$ is the volume bounded by $S$, which we take to
completely envelop one of the electrodes, as shown in the figure.
From equation \ref{didt} we see that, 
under the conditions expressed by equation \ref{rhodd},
\begin{equation}
\frac{dI_{S}(t)}{dt} = 0
\label{ic}
\end{equation}
for any choice of $S$. 
Equation \ref{ic} by itself says nothing about the actual value of
$I_{S}(t)$ or about the dependence of $I_{S}(t)$ on $S$. 
Equation \ref{ic} simply describes a {\em generic} type of conducting state 
in which matter (in this case, electrons) is being transferred
from region to region {\em at a steady rate}. We 
{\em define} such a state as a {\em true} global steady state.
We have seen that such states are minima of the quantity $A$
in equation \ref{a}. By using the above procedure
in the limit $t_{2}\rightarrow t_{1}$,
we obtain a definition of a type of conducting state that we 
call a true {\em instantaneous} global steady state.

The minimisation procedure applied to $A$ above treats $\rho({\bf r},t)$ as
a freely adjustible function of time, without regard for the actual physical dynamical
laws governing the electrons. We did invoke the continuity equation, which is a 
physical equation, but we did so in interpreting the {\em results} of the
minimisation of $A$, not in the minimisation itself. 

A real gas is governed by microscopic dynamical equations, which
may not permit the system to attain a true instantaneous global steady state,
as defined above. Our next task, therefore, is to seek the dynamical state {\em closest} to 
a true instantaneous global steady state, permitted by the laws of motion
that govern our system. We choose quantum mechanics as the dynamical 
law in question. Our electrons then are described by 
a many-body state vector $|\psi(t)\rangle$, governed
by the time-dependent Schr{\" o}dinger equation
\begin{equation}
i\hbar \,\frac{d|\psi(t)\rangle}{dt} = H \,|\psi(t)\rangle\,\,\,{\rm with}
\,\,\,|\psi(0)\rangle = |\psi_{0}\rangle
\end{equation}
where $H$ is the many-body electron Hamiltonian.
For the moment, we regard the initial condition $|\psi(0)\rangle = |\psi_{0}\rangle$
as a parameter. 

The electron density and current density are given by
\begin{eqnarray}
&& \rho({\bf r},t) = \langle\psi(t)|{\hat \rho}({\bf r})|\psi(t)\rangle\\
&& {\bf j}({\bf r},t) = \langle\psi(t)|{\hat {\bf j}}({\bf r})|\psi(t)\rangle\label{j}
\end{eqnarray}
where ${\hat \rho}({\bf r})$ and ${\hat {\bf j}}({\bf r})$ are the many-body
electron number-density and current-density operators, respectively.
The time-dependent Schr{\" o}dinger equation -- our chosen
dynamics -- guarantees equation \ref{cont} \cite{comment2}. We now
build this dynamical property of the system into the functional
that we have chosen as a measure of how close our system is to a steady state.
Substituting equation \ref{cont} into equation \ref{a}, in the
limit $t_{2}\rightarrow t_{1}$ we obtain the instantaneous functional, at a given time $t$, 
\begin{equation}
B[|\psi (t)\rangle] = \int_{{\rm all\,space}}\,d{\bf r} \,
\left( {\bf \nabla}\cdot {\bf j}({\bf r},t)\right)^{2} 
\label{b}
\end{equation}
with ${\bf j}({\bf r},t)$ given by equation \ref{j}.

We now define the instantaneous dynamical state closest to a true steady state by the following
variational procedure. Let the electron system have a given total energy
$E = \langle\psi(t)|H|\psi(t)\rangle$.
In our microcanonical picture, $E$ is a constant of the motion. We
select an arbitrary but definite surface $S$ of interest, and 
choose a value for the current $I_{S}$ across $S$. 
We write
\begin{equation}
|\psi(t)\rangle = \sum_{i}\,c_{i}\,|\psi_{i}\rangle\label{psi}
\end{equation}
where $\{|\psi_{i}\rangle\}$ are the many-body {\em bound} states of the 
system in figure 1, with eigenenergies $E_{i}$,
and $\{c_{i}\}$ are expansion coefficients. 
The reason for restricting the expansion to the bound part
of the spectrum of $H$ is that we do not wish to allow ionisation
of the electron-junction-electrode system. Ionisation would correspond to the 
escape, through the vacuum, of some finite electronic charge to infinity and, 
therefore, would not correspond to the experimental realisation of DC transport. 
We substitute the expansion in equation \ref{psi}
into equation \ref{j}, and thence into equation \ref{b}.
We then seek minima, with respect to $\{c_{i}\}$, of
\begin{eqnarray}
B[\{c_{i}\}] &&= \sum_{i,i',i'',i'''} \,c_{i}^{*}c_{i'}c^{*}_{i''}c_{i'''}\times\nonumber \\
&&\times \int_{{\rm all\,space}}\,d{\bf r} \,
({\bf \nabla}\cdot {\bf j}_{ii'}({\bf r}))\,
({\bf \nabla}\cdot {\bf j}_{i''i'''}({\bf r}))
\end{eqnarray}
where ${\bf j}_{ii'}({\bf r}) = 
\langle\psi_{i}|{\hat {\bf j}}({\bf r})|\psi_{i'}\rangle$,
subject to the constraints
\begin{eqnarray}
&&I_{S} = \sum_{i,i'} \,c_{i}^{*}c_{i'}\,
\int_{S} {\bf j}_{ii'}({\bf r})\cdot d{\bf S}\\
&&E = \sum_{i} \,c_{i}^{*}c_{i} E_{i}\\
&& \sum_{i} \,c_{i}^{*}c_{i}  = 1
\end{eqnarray}
Each solution for $|\psi\rangle = \sum_{i}\,c_{i}\,|\psi_{i}\rangle$ 
is an {\em instantaneous} many-body state, call it $|\psi (E,I_{S},t)\rangle$, 
containing only electrons bound within
the electrode-junction-electrode system in figure 1, that  
produces a given current $I_{S}$ and a given total energy $E$, 
while globally minimising the divergence of the current density.
The time $t$ is not part of the actual variational procedure
that generates the state $|\psi (E,I_{S},t)\rangle$. This
state is just a snapshot. We {\em assign} a time
$t$ to the snapshot, for the sake of being able to relate
the steady state in the snapshot to a definite initial condition, at an arbitrary 
but definite initial time $t = 0$.

We call the solution $|\psi (E,I_{S},t)\rangle$ 
an instantaneous {\em quasi} steady state: the best our
finite system can do to mimic a true instantaneous global steady state.
We may now explicitly write the initial condition required for the given 
instantaneous quasi steady state as
\begin{equation}
|\psi_{0}(E,I_{S})\rangle = e^{i H t/\hbar}\, |\psi(E,I_{S},t)\rangle
\end{equation}
This minimisation procedure may lead to more than one quasi steady state 
solution $|\psi (E,I_{S},t)\rangle$ for a given $E$ and $I_{S}$, 
in other words there may be different (in terms of charge and 
current densities) microscopic realisations 
of the {\it same} steady-state current. There may be combinations of $E$ and $I_{S}$,
for which no solution for $|\psi (E,I_{S},t)\rangle$ exists.
Finally, there may be initial conditions $|\psi_{0}\rangle$ that do not ever lead to a 
quasi steady state, i.e. that cannot be reached by back propagation
from any $|\psi (E,I_{S},t)\rangle$.

The variational nature of the quasi steady state enables us to draw
the following conclusion. At a quasi steady state
$|\psi (E,I_{S},t)\rangle$, $B$ is, by construction, stationary
against variations of $|\psi\rangle$ about $|\psi (E,I_{S},t)\rangle$,
compatible with the constraints. But $B$ is a measure of
the magnitude of the divergence of the current density, 
at least in a macroscopically averaged
sense. Thus, we may expect the quasi steady state flow pattern itself to be relatively
insensitive to variations about $|\psi (E,I_{S},t)\rangle$, at least on a
coarse-grained scale. 
However, after back-propagation to $t=0$, the corresponding spread of initial conditions, 
about $|\psi_{0}(E,I_{S})\rangle$, may contain large variations
in microscopic quantities such as the charge density. In other words,
there may be ``pockets'' of initial conditions (in Hilbert space), 
which we denote symbolically by $P_0(|\psi (E,I_{S},t)\rangle)$,
that differ in their microscopic properties but that produce the same, or
nearly the same, quasi steady state flow pattern at some later time $t$.
This conclusion supports the intuitive notion that the steady state 
should be relatively insensitive to the microscopic detail in the 
initial conditions. This conclusion also suggests a link with the notion of entropy.
The likelihood of a system with a given total energy $E$ attaining a steady state, 
$|\psi (E,I_{S},t)\rangle$,
with a given total current $I_{S}$ at time $t$, and the stability of this steady
state against small perturbations, is measured by the relative weight (in Hilbert space) 
of the ``pocket '' $P_0(|\psi (E,I_{S},t)\rangle)$, among all initial conditions
that lead to the current $I_{S}$ at $t$.

To develop this idea further, suppose that, for a given $E$ and $I_{S}$
we found several distinct quasi steady states $|\psi (E,I_{S},t)\rangle$.
We {\it postulate} that the quasi steady state 
$|\psi (E,I_{S},t)\rangle$, which would be observed, or preferentially
observed, in a macroscopic experiment on the system at time $t$, is that whose ``pocket''
$P_0(|\psi (E,I_{S},t)\rangle)$ has the largest statistical weight. 
In other words, the system is driven towards a specific microscopic quasi 
steady state at time $t$ by a  ``maximum-entropy principle'' where the 
``entropy'' measures the number of different initial conditions that 
realise the given steady state. The ``entropy'' introduced here both has a 
classical thermodynamic meaning and it contains the system dynamics through 
the minimisation procedure. The present maximum-entropy principle is a reinterpretation,
in the terms of the present microcanonical picture, of the 
approach used in reference \cite{ths} in the grand-canonical case,
in a mean-field one-electron picture.

There may exist quasi steady state solutions that have the same $E$ but
different $I_{S}$. The simplest interpretation of such solutions 
is that they represent different degrees of the discharge of the system. 
Thus, if we assign the same $t$ to them, their respective initial
conditions would correspond to different initial voltage drops in the
system. Alternatively, if we insist that their initial conditions have
the same, or comparable, voltage drops, then steady state solutions
with different $I_{S}$ would correspond to different $t$.
This does not mean, however, that steady states with the same
$E$ but with different $I_{S}$ are necessarily connected by a single
continuous evolutionary path: one solution may or may not
occur as a dynamical evolution of another. 
However, there may in fact be cases where solutions with the same 
$E$ but different $I_{S}$ describe steady states of different currents 
at the same time $t$ corresponding to the {\it same} voltage 
drop. Such instances correspond to chaotic transport and may occur in systems with 
intrinsic nonlinear dynamics \cite{zwolak}.

Finally, there is a further, highly speculative but intriguing, possibility. 
If we consider the quasi steady state solutions in an ensemble \{$I_{S}$\} 
of currents and form a linear combination of many-body wavefunctions out 
of their respective pockets of initial conditions, 
the system with this new initial condition could possibly evolve in time into the 
quasi steady state of yet {\it another} $I_S$, that does not belong to the original 
ensemble \{$I_{S}$\}. If this happens, then the system can fluctuate 
coherently between microscopic quasi steady states with {\it different} currents 
and steady state noise is produced. This additional noise has nothing to 
do with the ordinary (shot) noise due to 
charge quantization \cite{chen}; instead, it would be due to possible realisations of 
a steady state as a linear combination of microscopic steady states 
corresponding to different currents.

Let us now consider the quasi steady state from the point of
view of a practical measurement or a time-dependent calculation.
We let the system go from some initial state that we assume belongs
to a ``pocket'' $P_0(|\psi (E,I_{S},t)\rangle)$,  
such that, at some later time $t$, a quasi steady state with a total
current $I_{S}$, across a chosen surface $S$, is established. 
Let us then consider how we can define a conductance in this finite-system approach. 
Now that we have dispensed with the infinite reservoirs,
we may no longer appeal to B{\" u}ttiker's definition of conductance,
with respect to bulk electrochemical potentials of reservoirs in a
multiprobe measurement \cite{buttiker}. We thus fall back on Landauer's non-invasive 
definition of conductance, with respect to the electrostatic potential
drop in the system \cite{landauer}. The electrostatic potential $\phi ({\bf r},t)$,
subject to the boundary condition $\phi ({\bf r},t)\rightarrow 0$ as $r\rightarrow \infty$
appropriate to our isolated finite system, is a unique functional of the 
electron density $\rho({\bf r},t)$ and is thus unambiguously known
in the quasi steady state, or in any other state for that matter.
We take it as a physically plausible stipulation that, for large
enough electrodes, in a quasi steady state $\phi$ will tend (possibly
within microscopic Friedel-like oscillations) to 
well-defined values $\phi_{L}$ and $\phi_{R}$ in the interior
of the left and right electrodes of figure 1, respectively,
enabling us to define a potential difference $W = \phi_{L} - \phi_{R}$,
with respect to which we may then define conductance.

Let us now briefly consider processes that would 
help the system establish a steady state in the present phonon-free, microcanonical
picture. The obvious ones are electron-electron interactions.
Screening keeps the electron density 
macroscopically constant, a condition often referred to as charge neutrality.
This effective ``incompressibility'' of the electron gas in the metal
makes current flow somewhat analogous to water flow: local disturbances
in the density are not tolerated and heal fast. A further effect of electron-electron
interactions, and electronic $U$-processes in particular, is to 
produce relaxation of the total electron momentum in the electrodes, 
a quantity which is not a constant of the motion in our finite system.
We suggest, however, an additional intrinsic mechanism that facilitates
relaxation in the crucial region of the junction.
This mechanism is provided simply by the geometrical constriction experienced 
by electron wavepackets as they approach the nanojunction~\cite{payne}. This relaxation 
mechanism is due to the wave properties of the electron
wavefunctions and the resultant uncertainty principle, and has nothing to do 
with electron-electron interactions. Let us assume that the 
nanojunction has width $w$ and an electron wavepacket moves into it. 
The wavepacket has to adjust to the motion appropriate to the given junction geometry 
in a time $\Delta t \sim \hbar /\Delta E$, where $\Delta E$ is the typical energy 
spacing of lateral modes in the constriction. With $\Delta E\sim \pi^{2}\hbar^{2}/m_{{\rm e}}w^{2}$
we find $\Delta t \sim m_{{\rm e}}w^{2}/\pi^{2}\hbar$. 
For a nanojunction of width $w$ = 1 nm,
$\Delta t$ is of the order of 1 fs. In other words, even in the absence of 
inelastic effects, the mere presence of the nanojunction 
would contribute to relaxation of electron momentum. Thus, 
this effect would seem to suggest that even {\it without} electron-phonon or 
electron-electron inelastic scattering, a steady state could be reached in a nanoconstriction, 
with at most mean-field interactions. 

All of this analysis would be worthless if in order to do time-dependent
transport calculations one had to solve the many-body time-dependent Schr{\" o}dinger equation.
We conclude by showing that in the closed system of figure 1  
the true many-body total current is given {\it exactly} by the total current obtained with 
TDDFT~\cite{RungeGross}. This rigorous connection is 
independent of whether the system has reached a steady state or not.
We first note that for a given initial 
condition on the many-body wavefunction, a one-to-one correspondence 
between time-evolving charge density and external potential for the 
electron gas has been proven only when the density goes to zero at infinity (which is the case for a 
finite closed system with bound electrons)~\cite{RungeGross} or for an infinite but 
perfectly periodic solid~\cite{vignale}. Let us then assume that we have  
solved the time-dependent Kohn-Sham (KS) equations of TDDFT and obtained a 
set of KS time-dependent one-electron orbitals~\cite{RungeGross}. 
Referring to figure 1, we define the KS total current $I_{S}^{(KS)}$ by
\begin{equation}
I^{(KS)}_{S}(t)=\int_S {\bf j}^{(KS)}({\bf r},t)\cdot d{\bf S}=\int_V \,d{\bf r}\,{\bf \nabla}\cdot{\bf j}^{(KS)}({\bf r},t)
\end{equation}
where ${\bf j}^{(KS)}({\bf r},t)$ is the sum of expectation values of the one-electron current-density
operator in the populated KS orbitals. For densities that are non-interacting 
$v$-representable~\cite{leuwen}, the charge density of the true many-body system, $\rho({\bf r},t)$, is the 
same as the charge density obtained from the KS orbitals, $\rho^{(KS)}({\bf r},t)$.
Furthermore, $\rho^{(KS)}({\bf r},t)$ and ${\bf j}^{(KS)}({\bf r},t)$ satisfy the continuity
equation, just like the many-body density $\rho({\bf r},t)$ and current density 
${\bf j}({\bf r},t)$ in equation \ref{cont}. Since $\rho^{(KS)}({\bf r},t) =\rho({\bf r},t)$, 
we then have $ {\bf \nabla}\cdot{\bf j}^{(KS)}({\bf r},t) ={\bf \nabla}\cdot{\bf j}({\bf r},t)$
(even though ${\bf j}^{(KS)}({\bf r},t)$ and ${\bf j}({\bf r},t)$ need not be equal).
Hence,
\begin{equation}
I^{(KS)}_{S}(t)=\int_V \,d{\bf r}\,{\bf \nabla}\cdot{\bf j}^{(KS)}({\bf r},t)=
\int_V \,d{\bf r}\,{\bf \nabla}\cdot{\bf j}({\bf r},t) = I_{S}(t)
\end{equation}
where $I_{S}(t)$, once again, is the true total many-body electron current, 
across an arbitrary surface $S$.
The above proof is valid only for a {\it finite} system because, for an infinite system, 
we could not have made the transitions between surface and volume integrals
above. We note also that in practical calculations, due to the non-local nature of 
the exchange-correlation kernel in TDDFT, a formulation that relates the external potential 
directly to the current density may be numerically more efficient. 
(This formulation is known as time-dependent current-density functional theory~\cite{vk}.)

In  conclusion, we have analysed an alternative point of view of transport in nanoscale systems, 
in which two large but finite charged electrodes discharge across a nanoscale junction. This microcanonical 
formulation has key advantages. It permits the notion of a 
steady state to be expressed in variational form. This variational procedure suggests a link with the notion of 
entropy and a new source of steady state noise. It also allows one to show rigorously that 
the total current in a many-body system is given exactly 
by the corresponding quantity in TDDFT. This correspondence puts on rigorous 
footing the calculation of dynamical transport properties 
in nanoscale systems by the use of effective one-electron time-dependent 
Schr{\" o}dinger equations, without the 
need to implement scattering boundary conditions.
In particular, it allows the investigation of the onset, microscopic nature and dependence on 
the initial conditions of steady states, and may help tackle
open questions about the assumptions of the standard static approach to
steady-state conduction.
\newpage
\noindent
{\bf Acknowledgements}\\
\noindent
We thank G. Vignale, E. K. U. Gross, K. Burke, A. P. Horsfield and D. R. Bowler
for illuminating discussions. MD acknowledges support from the NSF 
Grant No. DMR-01-33075.


\begin{thebibliography}{99}
\bibitem{prec} Here we consider DC bias conditions. 
Similar considerations, however, apply to the AC case, 
that is, periodic time-varying disturbances of the charge.
\bibitem{comment0} If we do allow dissipative effects \cite{horsfield}, 
then a finite system is expected to reach a truly time-indepedent steady state, in the
course of its evolution. If the system is infinite, then this could be a current-carrying
steady state.
\bibitem{comment1} Here we assume, as we will do again later, that all electrons 
are bound within the electrode-junction-electrode system, so that the charge and 
current densities are exponentially small into the vacuum, at all times.
\bibitem{comment2} In fact, the current density operator in quantum mechanics is
specifically constructed so that the continuity equation is obeyed, but, with quantum
mechanics already in place, this does not affect our logic here.
\end{thebibliography}
\end{document}